\documentclass[prb,twocolumn,amssymb,amsfonts,amsmath]{revtex4-1}
\usepackage[a4paper,textwidth=17cm,textheight=25cm]{geometry}
\usepackage{epsfig}
\usepackage{graphicx}
\usepackage{hyperref}
\begin{document}

\title {Life under a black sun}
 \author{Tom\'{a}\v{s} Opatrn\'{y} and Luk\'{a}\v{s} Richterek}
 \email{tomas.opatrny@upol.cz}
 \email{lukas.richterek@upol.cz}
 \affiliation{Faculty of Science, Palack\'{y} University, 17. Listopadu 12,
 77146 Olomouc, Czech Republic}
 \author{Pavel Bakala}
 \email{pavel.bakala@fpf.slu.cz}
 \affiliation{Institute of Physics, Faculty of Philosophy and Science, Silesian University in Opava, Bezru\v{c}ovo n\'{a}m. 13, CZ-74601 Opava, Czech Republic}
\date{\today }
\begin{abstract}
Life is dependent on the income of energy with low entropy and the disposal of energy with high entropy. On  Earth, the low-entropy energy is provided by solar radiation and the high-entropy energy is disposed as infrared radiation emitted into the cold space. Here we turn the situation around and assume cosmic background radiation as the low-entropy source of energy for a planet orbiting a black hole into which the high-entropy energy is disposed. We estimate the power that can be produced by thermodynamic processes on such a planet, with a particular interest in planets orbiting a fast rotating Kerr black hole as in the science fiction movie {\em Interstellar}. We also briefly discuss a reverse Dyson sphere absorbing cosmic background radiation from the outside and dumping waste energy to a black hole inside.
\end{abstract}
\maketitle

\section{Introduction}
Life on Earth is possible thanks to the hot Sun and the cold sky. Their temperature difference makes it possible to drive processes far from thermodynamic equilibrium by increasing the entropy elsewhere in the Universe. Absorbing photons from the Sun at $\sim$\,6000\,K and emitting about 20~times more photons at $\sim$\,300\,K to the cold sky makes the entropy balance sufficient to sustain complex processes in which entropy locally drops. As explained by Erwin Schr\"{o}dinger in his book ``What is life?'',\cite{Schrod} organisms feed on negative entropy. The hot Sun and cold skies provide the Earth with a great deal of negative entropy: in this way, the Earth produces $\sim 5\times 10^{14}$ J/K of entropy each second.\cite{EarthEnergy}

\begin{figure}
\centerline{\epsfig{file=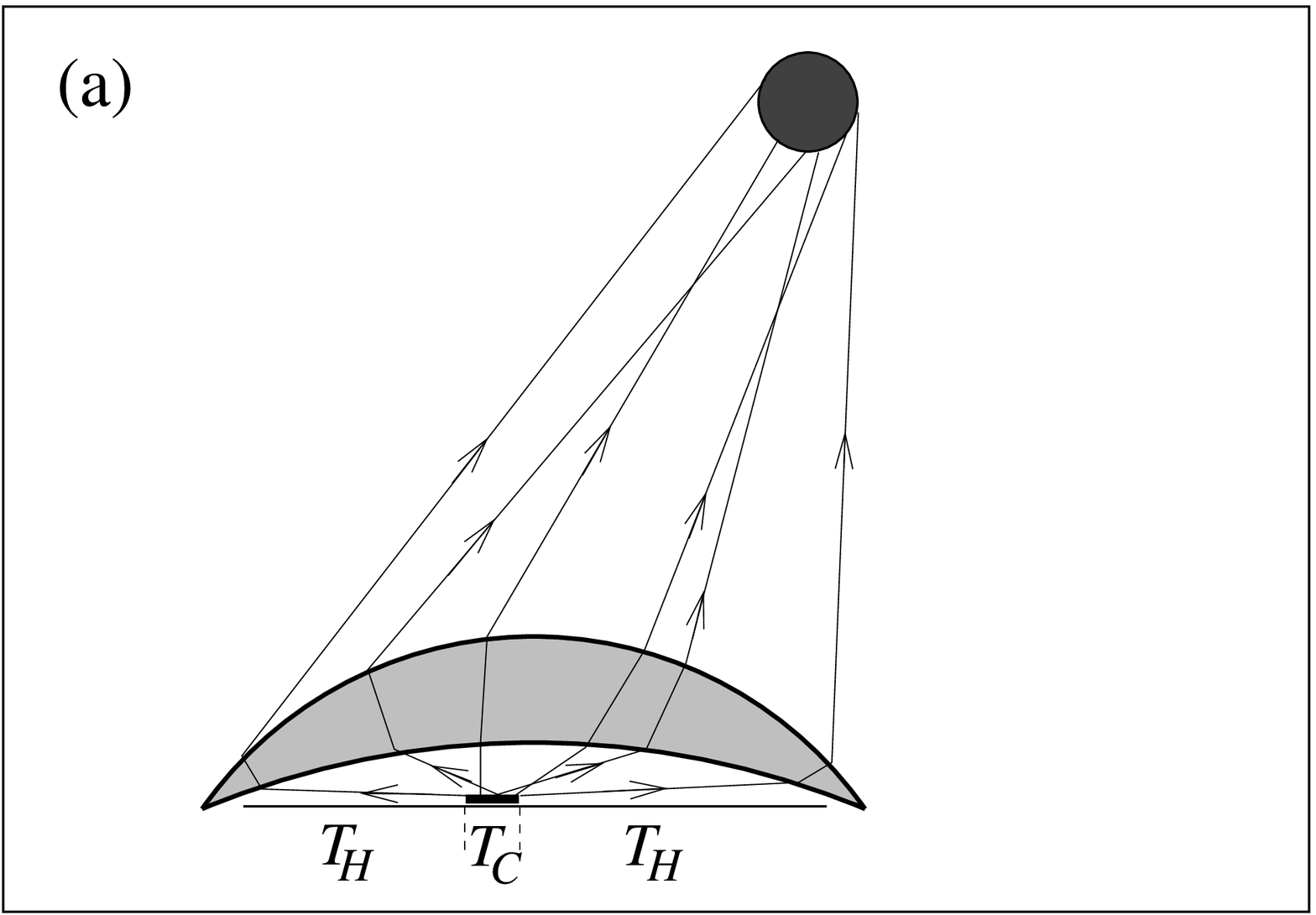,width=0.7\linewidth}}
\centerline{\epsfig{file=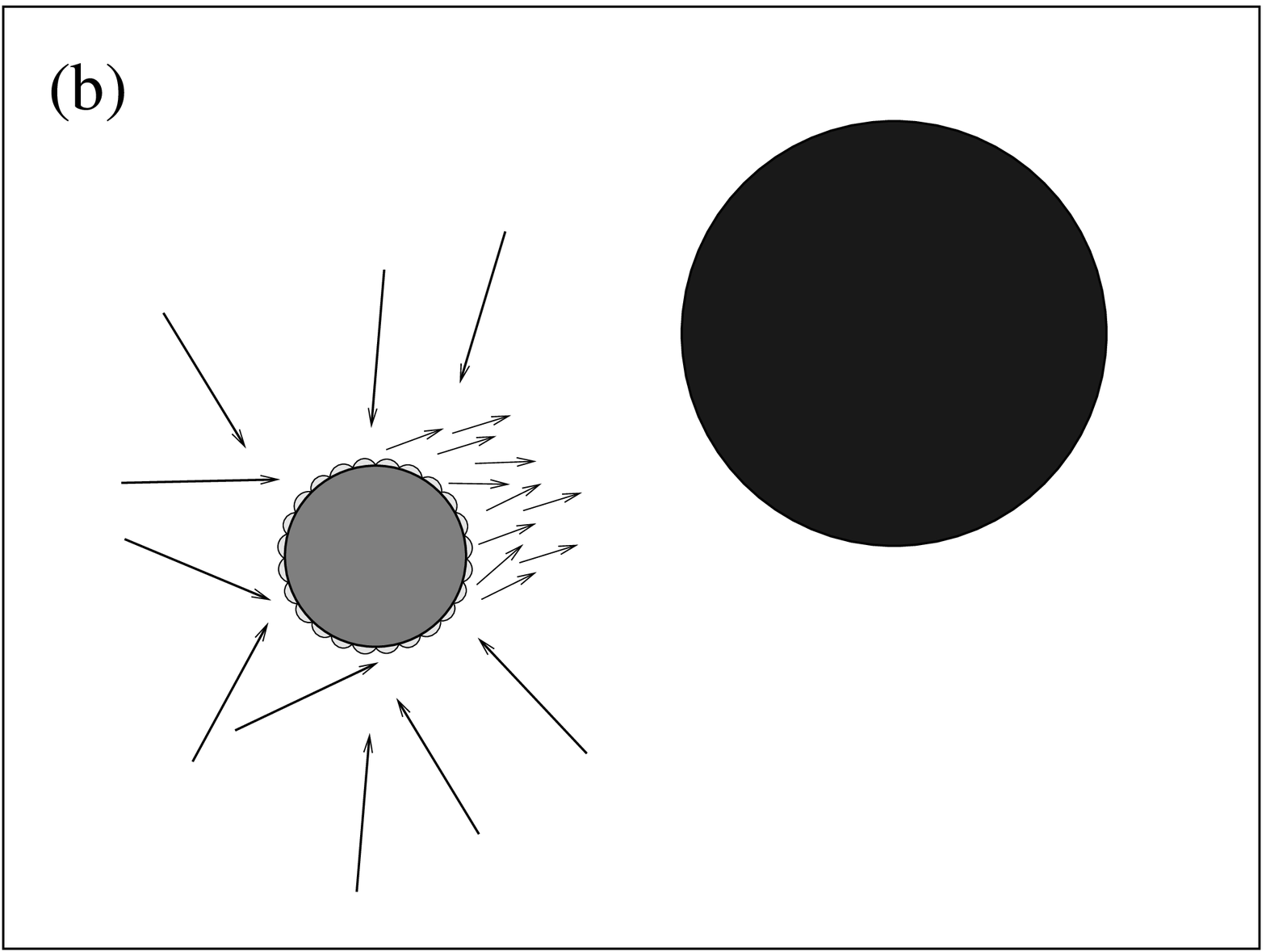,width=0.7\linewidth}}
\caption{\label{f-optics}
$(a)$ Projection system for interaction of a Lambertian radiator with the black hole. The arrows represent radiation from the cold surface at temperature $T_c$ directed to the black hole. The remaining surface at temperature $T_H$ interacts by radiation (not shown) with the hot sky.   $(b)$ Scheme of the thermodynamic system: the planet covered with the light concentration systems of the above picture accepts high energy photons (long arrows) from the space and sends low energy photons (short arrows) to the black hole.
}
\end{figure}

Here we play with the idea of a world upside down: the sun is cold and skies are ``hot''. Let us imagine a planet orbiting a black hole in a universe filled with background radiation. The inhabitants accept low-entropy energy from the sky and dispose waste heat to the black hole. The protagonists of the recent movie {\em Intersellar} who want to colonize a planet orbiting a supermassive black hole {\em Gargantua} might find these results vital (for physical details of their trip and destination we recommend the book ``The Science of Interstellar''\cite{Thorne} by Kip Thorne, the scientific consultant and an executive producer of the film). 

Apart from the pedagogical value of simple thermodynamic exercises, these speculations might be relevant in the distant future when stars exhaust their nuclear fuel and die, and black holes may become dominant constituents of the entropy production processes.\cite{Frautschi,Krauss} This kind of energetics might be useful until the expansion of the Universe cools the cosmic background radiation below the temperature of the black hole Hawking radiation. After that, the black hole becomes a net radiator and thus the nearby inhabitants might again live under a ``hot'' sun and cold sky. Recently, we discussed in this journal a 
mechanism of extracting mechanical work of radiating black holes;\cite{Opatrny-Richterek} however, the focus here is on the not so distant future.

The considerations presented here can also be relevant for the discussion of the early stages of the Universe: recently Loeb\cite{Loeb} suggested that a habitable epoch occurred when the Universe was about 15 million years old and the background radiation had the temperature of 273--300 K, allowing for rocky planets with liquid water chemistry on their surfaces. This suggestion was covered by the Nature magazine\cite{Merali} mentioning also some criticism pointing out that the cold sky is thermodynamically as important for life as the hot Sun. Although there is no doubt that some source of negative entropy is necessary for life, the question remains what could be the source. The black-hole sun is an option to be discussed here.

The paper is organized as follows: In Sec.~\ref{Sec:Radiation} general formulas for converting the incoming background  radiation energy into useful work are discussed, in Sec.~\ref{Sec:regimes} two special regimes for the solid angle of the incoming radiation---large and small are studied, in Sec.~\ref{Sec:circular} the results for anisotropic radiation due to fast orbiting close to the black hole are presented, in Sec.~\ref{Sec:Dyson} a~Dyson sphere encapsulating a black hole is considered, and in Sec.~\ref{Sec:Conclusion} the conclusions are provided.  Details of the computations of the temperature map of the sky for an observer orbiting a black hole are given in appendixes.

\section{Radiation heat exchange}
\label{Sec:Radiation} 
\subsection{Sky projection and etendue conservation}
\label{Sec:Etendue} 
Let us assume that heat exchange of the planet can occur radiatively by surface $S$. The celestial sphere\cite{Celestial} 
is divided into two parts: one is ``hot'' at temperature $T_1$ and the other cold, for simplicity at temperature of absolute zero. 
The zero-temperature black hole is a good approximation if Hawking radiation\cite{Hawking76,Hawking77} is negligible.
Let the solid angles spanned by these two parts be $\Omega_H$ and  $\Omega_C$, with    $\Omega_H+ \Omega_C=4\pi$. To allow for the most efficient thermal energy exchange of the planet with the sky, we assume the whole planet to be covered with light concentration systems that project the celestial sphere into its images. We assume the numerical aperture of these devices approaches 1, which means that each point of the image is uniformly illuminated from the $2\pi$ solid angle by rays coming from the corresponding point of the object (see Fig.~\ref{f-optics}). This means that each part of the image can interact as a Lambertian radiator with the corresponding part of the sky (the defining feature of the Lambertian radiator is that its surface has the same radiance when viewed from any angle).
In this way, all rays coming from the hot part of the sky are projected to one part of the surface of the planet and rays from the cold part to another. Let us denote the areas of these surfaces $S_H$ and $S_C$ with   $S_H+S_C = S$. 

For the bundle of all possible rays going through some area of the system we can introduce a quantity called etendue (or \'etendue) which is equal to the area  multiplied by the solid angle occupied by all the ray directions. Generally, etendue characterizes how the light is spread in area and angle,  and being multiplied by radiance, it gives the radiation power.
A general theorem of ray optics says that the etendue for propagating light cannot decrease (see, e.g., Ref. \onlinecite{Markvart}). 
For a part of a beam propagating in directions within a solid angle $\delta \Omega$ across a surface element $\delta A$, the element of etendue is defined as $\delta {\cal E} = n^2 \cos{\theta} \delta \Omega \delta A$, where $n$ is the refraction index of the medium and $\theta$ is the angle between the direction of propagation and the normal to $\delta A$. If the light propagates through non-absorbing non-scattering environment (such as through our idealized optical systems), the etendue is conserved. This is analogous to the phase-space conservation in conservative systems according to the Liouville theorem. For our scheme the consequence of etendue conservation is that 
\begin{eqnarray}
\frac{S_H}{S_C} = \frac{\Omega_H}{\Omega_C}.  
\label{OmegaS}
\end{eqnarray}
This means that the total surface of the planet can be divided in two parts
$S_H$ and $S_C$ serving as the hot and cold terminals of heat engines, their proportion being equal to the proportion of the hot and cold parts of the sky.

Without the idealization leading to Eq.~(\ref{OmegaS}), one would have to take into account the fact that the cold terminal also interacts with objects at higher temperature which would decrease the efficiency of work production. Let us note that in the photovoltaic industry one of the goals is the construction of light concentrators approaching the etendue limit (see, e.g., Ref. \onlinecite{Winston}), so we assume that the advanced civilization inhabiting the planet under study has reached this goal.

\subsection{Temperature optimization for given $S_H$ and~$S_C$}
\label{Sec:temperature-optimization}
Let the surfaces of the heat exchangers $S_H$ and $S_C$ be at temperatures  $T_H$ and $T_C$ and let them serve as heat exchangers for a heat engine. The task is to find such $T_H$ and $T_C$ that the power of the heat engine is maximized. The procedure is analogous to the power optimization of irreversible engines as first studied by Novikov\cite{Novikov} and later independently 
in this journal 
by Curzon and Ahlborn\cite{Curzon} (for more general considerations see, e.g., Ref. \onlinecite{Bejan}). Our case is different in the temperature dependence of the thermal energy exchange rate ($\propto T^4$), and also in having fixed the ratio of the exchanger surfaces  $S_H/S_C$.

Thermal energy received by  $S_H$ from the hot part of the sky during a time interval $\Delta t$ is $Q_1 = \sigma S_H (T_1^4 - T_H^4)\Delta t$ and the waste energy sent to the cold part of the sky is $Q_2 = \sigma S_C T_C^4\Delta t$, where $\sigma$ is the Stefan-Boltzmann constant, $\sigma \approx 5.68\times 10^{-8}$\,W\,m$^{-2}$\,K$^{-4}$. Their difference can be converted into work $W=Q_1-Q_2$ provided that $Q_1/Q_2 = T_H/T_C$. A relation between $T_H$ and $T_C$ follows from this assumption,
\begin{eqnarray}
 T_C = \left[ \frac{S_H}{S_C}\left( \frac{T_1^4}{T_H^4} - 1 \right) \right]^{1/3} T_H ,
\end{eqnarray}
and the average power $P=W/\Delta t$ can be expressed as
\begin{eqnarray}
\label{eq-work}
 P &=& \sigma S_H \\
& & \times \left[T_1^4 - T_H^4 
- \left( \frac{S_H}{S_C}\right)^{1/3}\left( \frac{T_1^4}{T_H^4} - 1 \right)^{4/3} T_H^4 \right]  .
\nonumber
\end{eqnarray}
Assuming that $S_H$,  $S_C$, and $T_1$ are fixed, one can find such $T_H$ that the power of the engine is maximized. Setting the derivative equal to zero, $dP/dT_H = 0$, one finds that 
\begin{eqnarray}
T_H &=& u^{1/4}T_1, \\
T_C &=& \left( \frac{S_H}{S_C}\right)^{1/3} \frac{(1-u)^{1/3}}{u^{1/12}} T_1,
\end{eqnarray} 
where $u$ solves the equation
\begin{eqnarray}
 \frac{27}{q} u^4 - 18 u^2 - 8 u -1 = 0
\label{Eq-u}
\end{eqnarray}
in the interval $0 < u < 1$, and 
\begin{eqnarray}
q\equiv \frac{S_H}{S}=\frac{\Omega_H}{4\pi}
\end{eqnarray}
is the ``hot'' fraction of the sky.
The resulting power is then
\begin{eqnarray}
P_{\rm max} &=& \eta q \sigma S T_1^4 , 
\label{Pmax}
\end{eqnarray}
where
 $q \sigma S T_1^4$ is the total power of the incoming radiation, and 
\begin{eqnarray}
\label{eq:eta}
\eta \equiv 1-u - \left(\frac{q}{1-q} \right)^{1/3}u \left( \frac{1-u}{u} \right)^{4/3} 
\end{eqnarray}
is the efficiency with which the incoming radiation can be converted into useful work.

\begin{figure}
\centerline{\epsfig{file=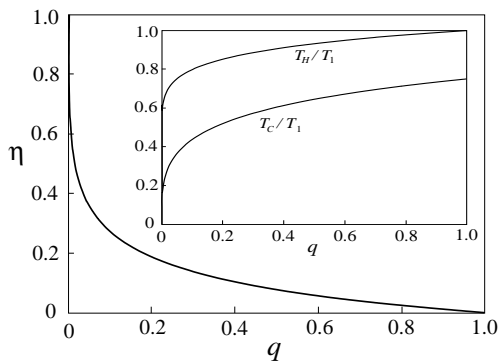,width=0.8\linewidth}}
\caption{\label{fbs1}
Efficiency $\eta$ in dependence on the hot area fraction $q$ for optimized power according to  Eq.~(\ref{eq:eta}). Inset: temperatures of the hot and cold surfaces in dependence on $q$.
}
\end{figure}

Even though Eq.~(\ref{Eq-u}) as a quartic equation has an explicit solution, we solve it numerically and the result is used to find the dependence of the working temperatures $T_H$, $T_C$ and of the efficiency $\eta$ on the hot-sky fraction $q$, as shown in  Fig.~\ref{fbs1}.

\section{Special operating regimes}
\label{Sec:regimes}
\subsection{Small heating area}
\label{Sec:small-heating}
Assuming  $S_H \ll S_C$, or  $q\ll 1$,  Eq.~(\ref{Eq-u}) yields $u\approx 3^{-3/4}q^{1/4}$ and the temperatures approach zero as 
\begin{eqnarray}
T_H &\approx& 3^{-3/16}q ^{1/16} T_1, \\
T_C &\approx& 3^{1/16}q^{5/16} T_1
\end{eqnarray}
(see Fig.~\ref{fbs1}), and  the efficiency $\eta$ of the process approach 1 as
\begin{eqnarray}
\eta \approx 1- \left(3q \right)^{1/4} . 
\end{eqnarray}

Rather than to a ``standard'' black hole, this limit corresponds to a distant star illuminating a planet in an otherwise empty universe (the relevance to a less conventional black hole situation will be discussed later in Sec.~\ref{Sec:Kerr}). The results can be applied to estimate Earth's energy income from our Sun. In this case  $q\approx 5.4\times 10^{-6}$ and $T_1=5778$\,K, which leads to working temperatures $T_H=$ 0.385 $T_1 \approx 2,200$ K and $T_C=$ 0.0239 $T_1 \approx 138$ K, yielding the efficiency 92 \%, and maximum available power $P \approx 1.603\times 10^{17}$ W.

\subsection{Large heating area}
\label{Subsec-bigHot}
In the opposite limit  $S_H \to S$, on solving Eq.~(\ref{Eq-u}) for $1-q \ll 1$ we find
\begin{eqnarray}
u \approx 1-\frac{3^3}{2^6}\left( 1-q \right)
\end{eqnarray}
which leads to
\begin{eqnarray}
T_{H} &\approx& \left[ 1- \frac{3^3}{2^8} (1-q) \right] T_1, \\
T_{C} &\approx& \frac{3}{4} \left[ 1- \frac{3^4}{2^8} (1-q) \right] T_1, \\
\eta  &\approx& \frac{3^3}{2^8}(1-q).
\end{eqnarray}
Using these expressions in (\ref{Pmax}), we find the power of maximum work generation
\begin{eqnarray}
P_{\rm max} \approx \frac{3^3}{2^8}\sigma S_C T_1^4.
\label{Power}
\end{eqnarray}

As an illustration, let us assume a planet of the Earth's size 
orbiting a black hole whose angular size as seen from the planet equals to the angular size of Sun as seen from Earth. This gives $q\approx 1-5.4\times 10^{-6}$ and $\eta \approx 5.7\times 10^{-7}$.
Thus, $S_C \approx 2,760$\,km$^2$, so that the planet would radiate its waste thermal energy from an area comparable to Rhode Island.  
If the background radiation is at  room temperature $T_1=$\,300\,K as in the  15~million years old universe assumed in Ref. \onlinecite{Loeb}, then $T_C \approx  225$\,K, i.e. about $-48^{\circ}$C. From Eq.~(\ref{Power}) we find that useful work could then be obtained with power $P_{\rm max} \approx 130$\,GW. This is two orders of magnitude below the present world energy consumption, and six orders of magnitude below the power presently supplied to Earth by Sun. The background radiation of today’s skies is two orders of magnitude colder at $T_1 = 2.725\,K$, which results in even less available useful power $P \approx 910\,W$.

\begin{figure}
\epsfig{file=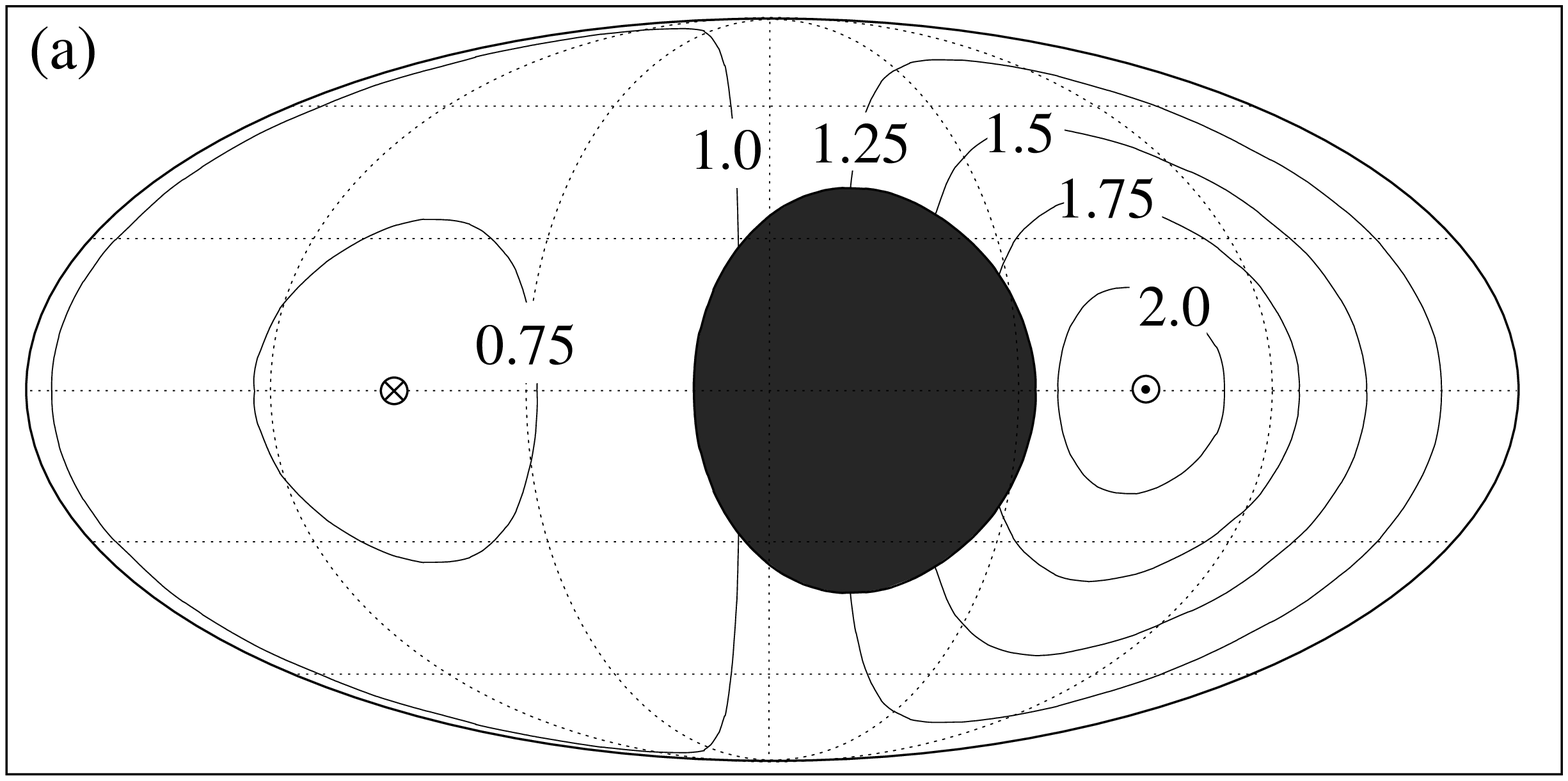,width=0.7\linewidth}
\epsfig{file=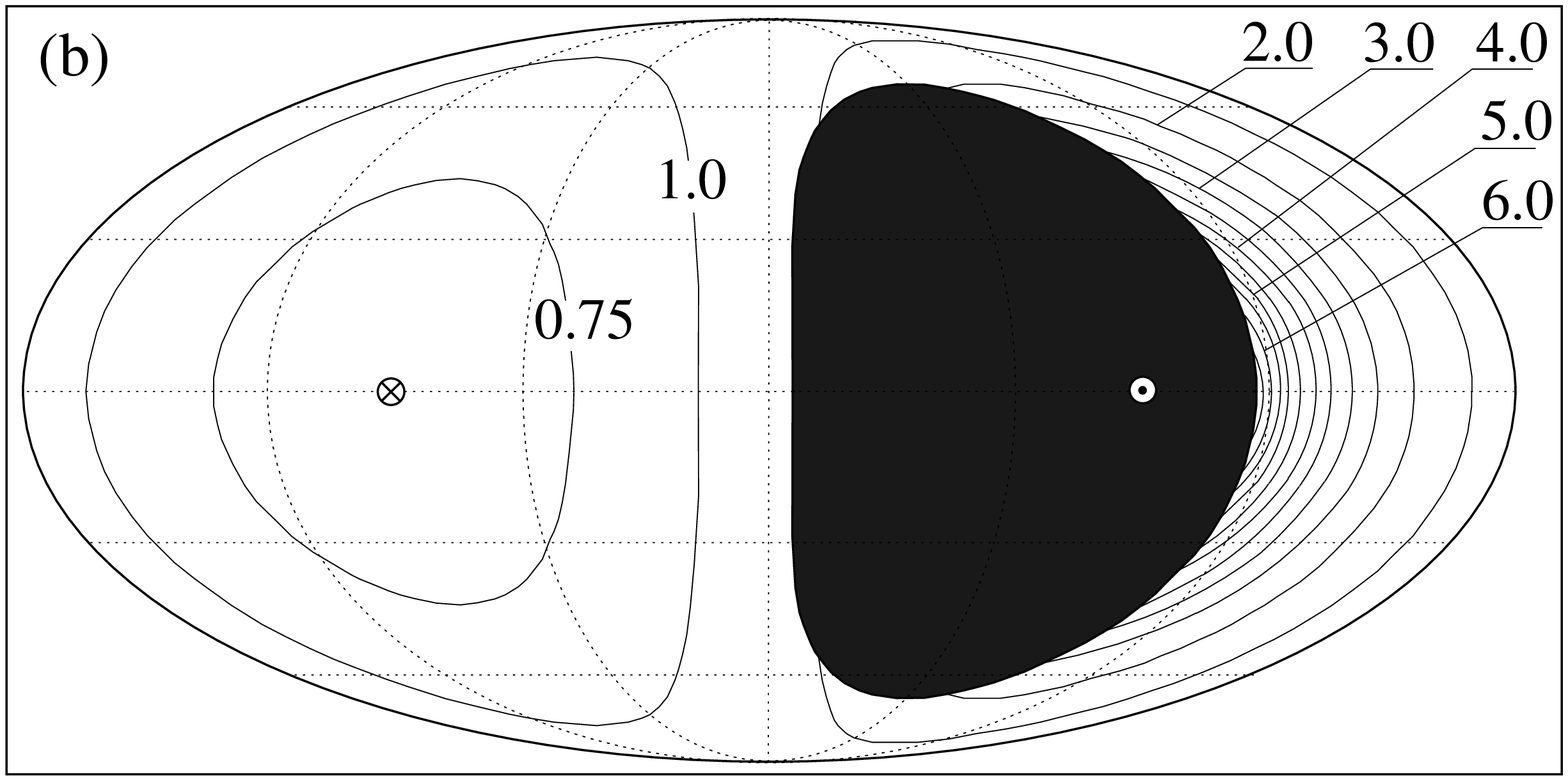,width=0.7\linewidth}
\epsfig{file=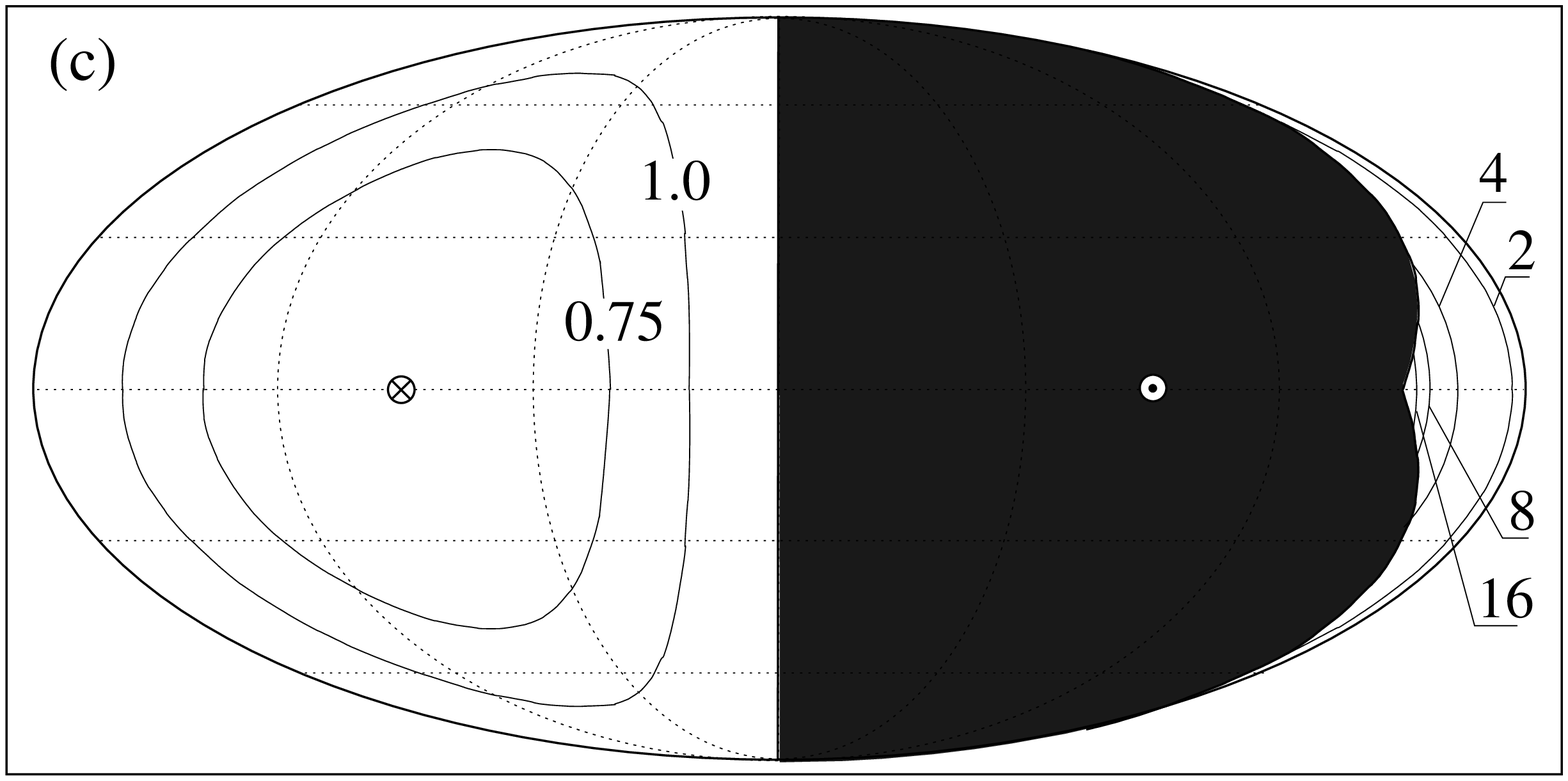,width=0.7\linewidth}
\caption{\label{fig:obloha}
Mollweide projection of the sky of an observer orbiting a black hole. The dark area is the shadow of the black hole, the full lines are contour lines of constant temperature, where the number indicates relative temperature shift with respect to the temperature of background radiation measured by a distant observer. The symbols $\odot$ and $\otimes$ indicate the direction to which and from which the observer is moving, respectively. The dotted lines are parallels of latitude $\theta = 0, \pm 30^{\circ}, \pm 60^{\circ}$ and meridians of longitude $\phi = 0, \pm 60^{\circ}, \pm 120^{\circ}$.
(a) A nonrotating black hole,  orbit at $r=6 GM/c^2$, (b) a rotating black hole
with rotation parameter $a = 1-1.3\times 10^{-14}$ and orbit radius $r= 2.2GM/c^2$, 
(c) a rotating black hole
with $a = 1-1.3\times 10^{-14}$ and $r= 1.0000379 GM/c^2$.}
\end{figure}

\section{Circular motion close to the black hole}
\label{Sec:circular}
\begin{figure}
\epsfig{file=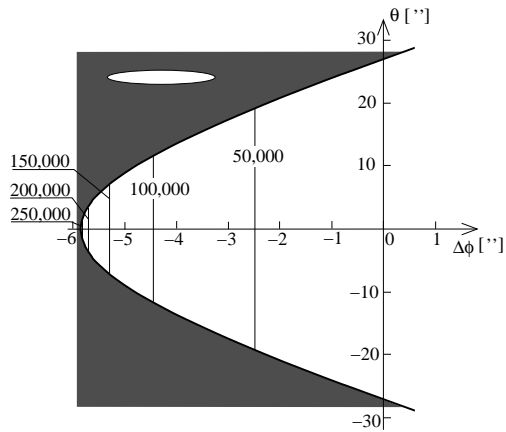,width=0.8\linewidth}
\caption{\label{fig:vecko1}
Detail of the shadow boundary of a black hole for the case of Fig.~\ref{fig:obloha}(c) near $\theta=0$ and $\phi=150^{\circ}$, where $\Delta \phi \equiv \phi-150^{\circ}$. The contour lines of constant temperature are shown, the number indicating relative temperature shift with respect to the temperature of background radiation measured by a distant observer. The light spot on the upper left shows the angular size of Neptune as seen from the Earth, for comparison (the shape is deformed due to different scales of $\theta$ and $\Delta \phi$).}
\end{figure}

Since $\eta$ increases with decreasing $q$, to get as much power as possible, we should place the orbit as close to the black hole as possible. 
However, things get complicated there: the observer moves fast and relativistic effects become important. The absorbed radiation is Doppler-shifted, as well as blue-shifted by falling to the black hole vicinity.
The influence of the gravitational blue-shift on the cosmic background radiation for an observer close to the event horizon of the Schwarzschild (i.e., nonrotating) black hole was recently studied in~Ref.~\onlinecite{Abramowicz:2014}.  Here we consider an observer located on the planet orbiting the black hole, both  Schwarzschild and Kerr (i.e., rotating).   
As a result of the motion, the sky ceases to be isothermal: the radiation comes hotter from the direction where the observer moves and colder from the rear. To get the precise picture of the sky,  computations outlined in Appendixes~\ref{Sec:analyticschw} and \ref{Sec:Computation} have been applied; the results are in Figs.~\ref{fig:obloha} and \ref{fig:vecko1}. Having the temperature map of the sky, one can  use segments of the sky with different temperature as heaters and apply another optimization to allocate to them various parts of the cooling area so as to get maximum power.

\subsection{Schwarzschild  black hole}
\label{sec:schwblackh}
We first consider motion along a timelike circular geodesic around a static, spherically symmetric black hole of a mass $M$ and Schwarzschild radius $R_{S}=2GM/c^2$, where $G=6.67\times10^{-11}$\,N\,m$^{2}$\,kg$^{-2}$ is the gravitational constant and $c=3\times10^{8}$\,m/s is the speed of light. Stable bound circular orbits of radius $r$ exist for $r\geq 3R_{S} = 6GM/c^2$ (see, e.g., Refs. \onlinecite{Raine:2009} and \onlinecite{Hartle:2003}). In order to obtain the maximum power, the innermost stable circular orbit at  $r = 6GM/c^2$ has been chosen. The resulting picture of the sky is in Fig.~\ref{fig:obloha}(a). The shadow of the black hole covers 12.2\,\% of the sky, i.e., $q=0.878$. If the temperature shifts were disregarded, Eqs.~(\ref{Eq-u}), (\ref{Pmax}), and (\ref{eq:eta}) would yield $\eta \approx 1.39$\,\% and $P_{\rm max}\approx 0.012\sigma S T_1^4$. For an Earth-size planet and background radiation at $T_1 = 2.725$\,K, this would be $P_{\rm max}\approx$\,19\,MW.

\begin{figure}
\epsfig{file=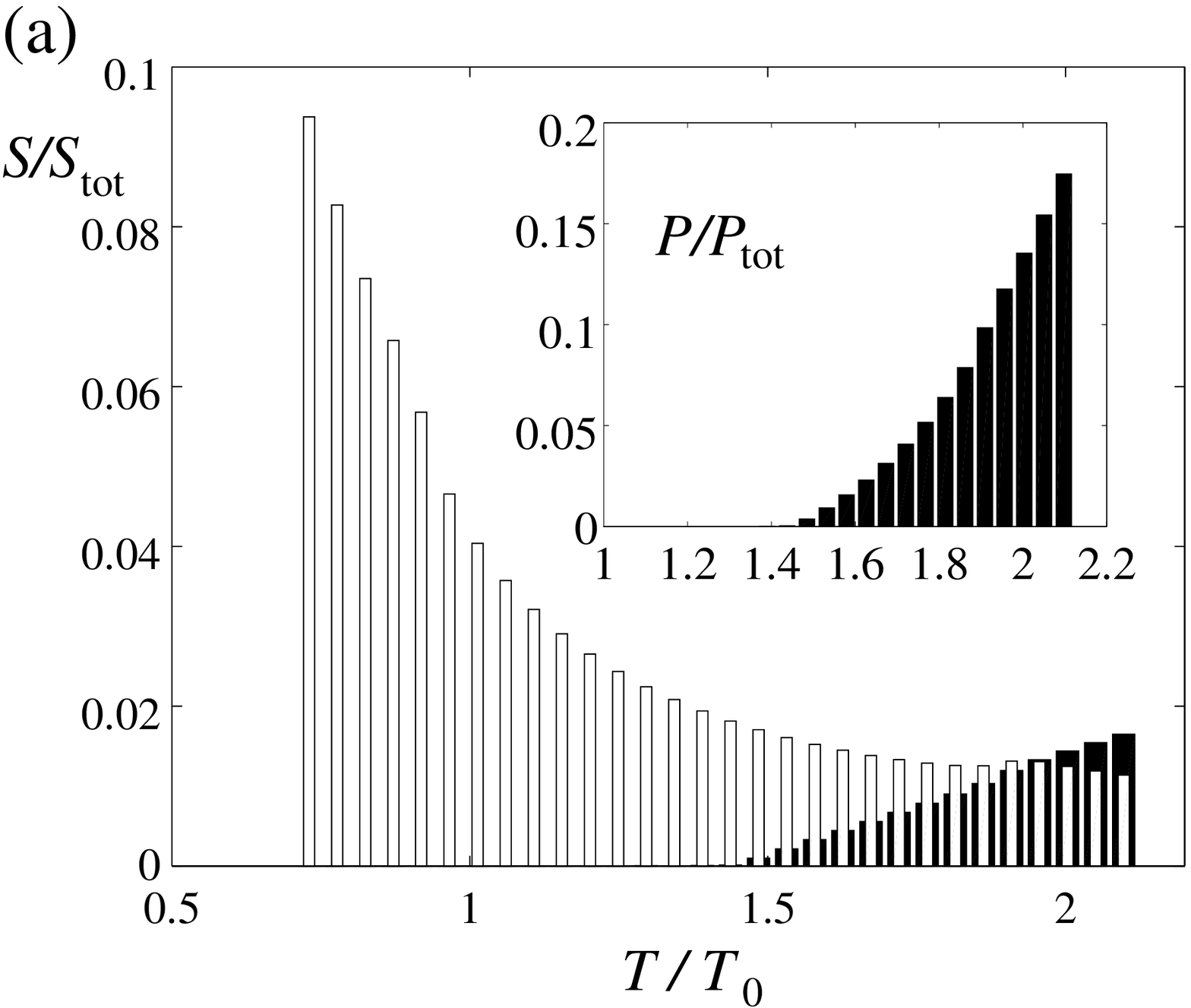,width=0.7\linewidth}
\epsfig{file=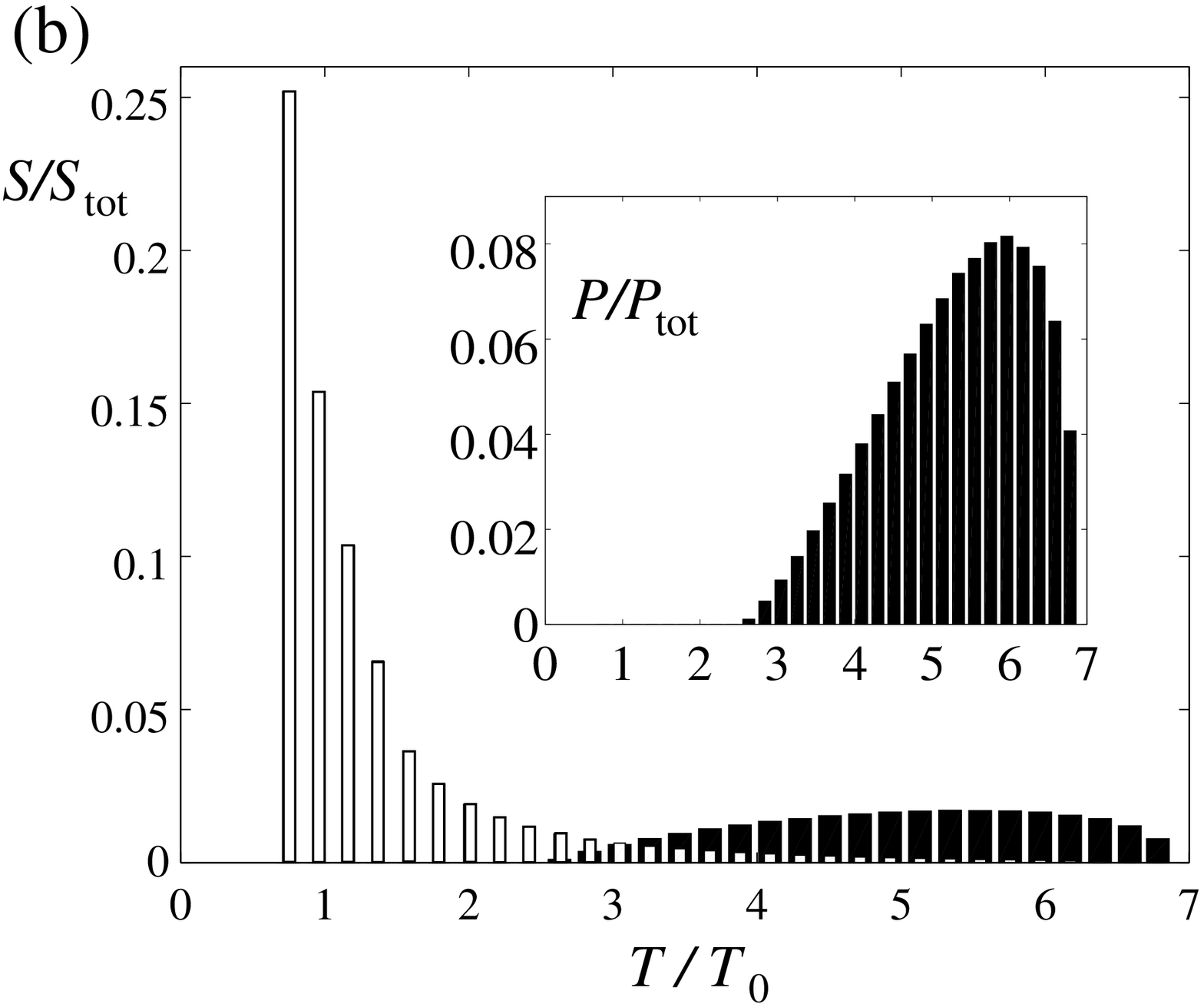,width=0.7\linewidth}
\caption{\label{fig:plochy}
Empty bars: fractions of observer's sky of different temperatures, where the ``hot'' sky was divided into 30 segments of equal temperature span. Full bars: fractions of observer's sky used as cold reservoirs for heat engines of upper temperature $T$. All bars sum up to 1, the empty bars sum up to the fraction of the radiating sky and the full bars sum up to the fraction covered by the shadow of black hole. Inset: fractions of the power of the heat engines with upper temperature $T$, the bars sum up to 1. 
(a) Schwarzschild black hole and orbit at  $r=6 GM/c^2$ as in  Fig.~\ref{fig:obloha}(a), (b) Kerr black hole with  $a = 1-1.3\times 10^{-14}$ 
and orbit at  $r=2.2 GM/c^2$  as in Fig.~\ref{fig:obloha}(b).}
\end{figure}

However, fast motion around the black hole makes parts of the sky warmer which brings some advantage. As seen in Fig.~\ref{fig:obloha}(a), the sky temperature is more than twice as large measured by the orbiting observer in the direction of motion compared to an observer at rest, far from the black hole (the maximum blue shift being $3/\sqrt{2} \approx$ 2.12; see~Appendix~\ref{Sec:analyticschw} for the derivation).  
Thus, the power available from this part of the sky can be more than $2^{4}$ times larger than without the blue shift. However, the hottest part of the sky is relatively small and the rest of the sky is colder. To find the total available power, we have divided the hot sky into 30 segments of equal temperature-span intervals. Each segment then serves as a heater of a separate heat engine. Since the power of each engine also depends on the cold area used for dumping its waste energy, one has to find the optimum allocation of the cold area to the individual engines to obtain a maximum total power. The result of the numerical optimization is shown in Fig.~\ref{fig:plochy}(a). The white bars are the areas of the hot segments which are given as the input data. The black bars are the areas of the corresponding cold segments obtained as the result of the optimization procedure.  
As can be seen, the optimum allocation assumes absorbing radiation from the 14 hottest segments of blueshifts above 1.4. Colder parts of the ``hot'' sky are not worth using as hot reservoirs.
The resulting power is then $P_{\rm max}\approx 0.126\sigma S T_1^4$, i.e., one order of magnitude larger than the above estimate based on disregarding the frequency shifts.  For an Earth-size planet and background radiation at $T_1 = 2.725$ K, this would be $P_{\rm max}\approx$ 200 MW, i.e., still rather low for a comfortable life of more than a few small towns. 

\subsection{Fast rotating Kerr black hole}
\label{Sec:Kerr}

How can we come closer to the black hole so as to increase the cold sky proportion? We were inspired by the movie {\em Interstellar} where the characters come close to a fast-rotating giant black hole named {\em Gargantua}. Rotating black holes have stable circular orbits closer than  $6GM/c^2$. For Gargantua, the rotation parameter $a$ was extraordinarily large, $a = 1-1.3\times 10^{-14}$, thus allowing the characters to enjoy extraordinarily strong relativistic effects close to the black hole.\cite{Thorne,Spin} We have computed the results for two special cases of the orbit radius,  $r = 2.2GM/c^2$ (Fig.~\ref{fig:obloha}(b) and  \ref{fig:plochy}(b)) and $r= 1.0000379 GM/c^2$  (Fig.~\ref{fig:obloha}(c) and  \ref{fig:vecko1}). The latter case corresponds to the orbit of {\em Miller's planet} where the characters of Interstellar spend three hours while 21 years pass on their base station which is sufficiently distant from the reach of Gargantua's gravitational time shift. 

We can see that the shadow of the black hole becomes deformed and covers a large part of the observer's sky, including the direction of motion of the planet. Nevertheless, the planet does not fall into the hole as it is dragged by its rotating gravitational field. The blue shift becomes much stronger, although in relatively small strips of the sky just above the shadow of the black hole in the direction of the planet motion. In the case of  $r = 2.2GM/c^2$ the black hole covers 26 \% of the sky and the maximum blue shift is 6.90. Upon using the same optimization procedure as in the preceding subsection, one finds the maximum attainable power to be $P_{\rm max}\approx 4.2\sigma S T_1^4$. For an Earth-size planet and background radiation at $T_1 = 2.725$ K this would be $P_{\rm max}\approx$ 6.7 GW, i.e., enough for a small country. 

The case of Miller's planet with  $r= 1.0000379 GM/c^2$ leads to extreme blue shifts, reaching up to 275,000. The black hole covers 40 \% of the sky and most of the radiation energy comes from a very narrow strip of a few arcseconds (see Fig.~\ref{fig:vecko1}): our numerical results show that 99\% of the energy comes from a strip of longitude span 2.3$^{\prime \prime}$ and latitude span  5.8$^{\prime \prime}$. This size is comparable, e.g., to the angular size of the planet Neptune as seen from Earth, with angular diameter $\approx  2.2^{\prime \prime}$. Thus, the good news is that Miller's planet enjoys small heating area regime (see Sec.~\ref{Sec:small-heating}), meaning that most of the incoming radiation energy can be converted into useful work. The bad news for the visiting astronauts is that it is too much energy: the incoming flux density (power per unit area perpendicular to the incoming radiation) is  $\Phi \approx 420$ kW/m$^2$, i.e., about 300 times bigger than the solar constant. This value can be used to find the equilibrium temperature of a planet radiating its energy as a black body, $T=\sqrt[4]{\Phi/(4\sigma)}\approx  890^{\circ}{\rm C}$. Thus, the tidal waves observed on the planet might be, e.g., of melted aluminum. Moreover, the astronauts would be grilled by extreme-UV radiation.

\section{Black hole Dyson sphere}
\label{Sec:Dyson} 

\begin{figure}
\centerline{\epsfig{file=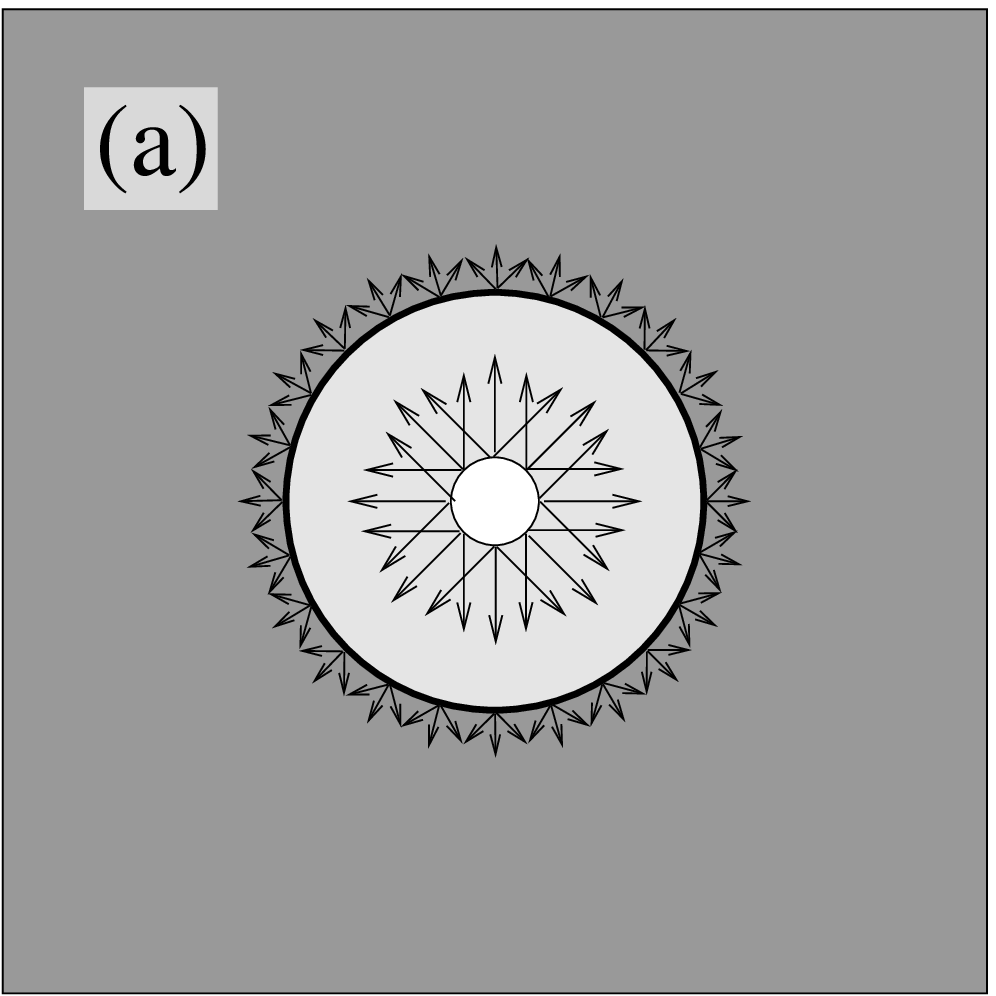,width=.5\linewidth}}
\centerline{\epsfig{file=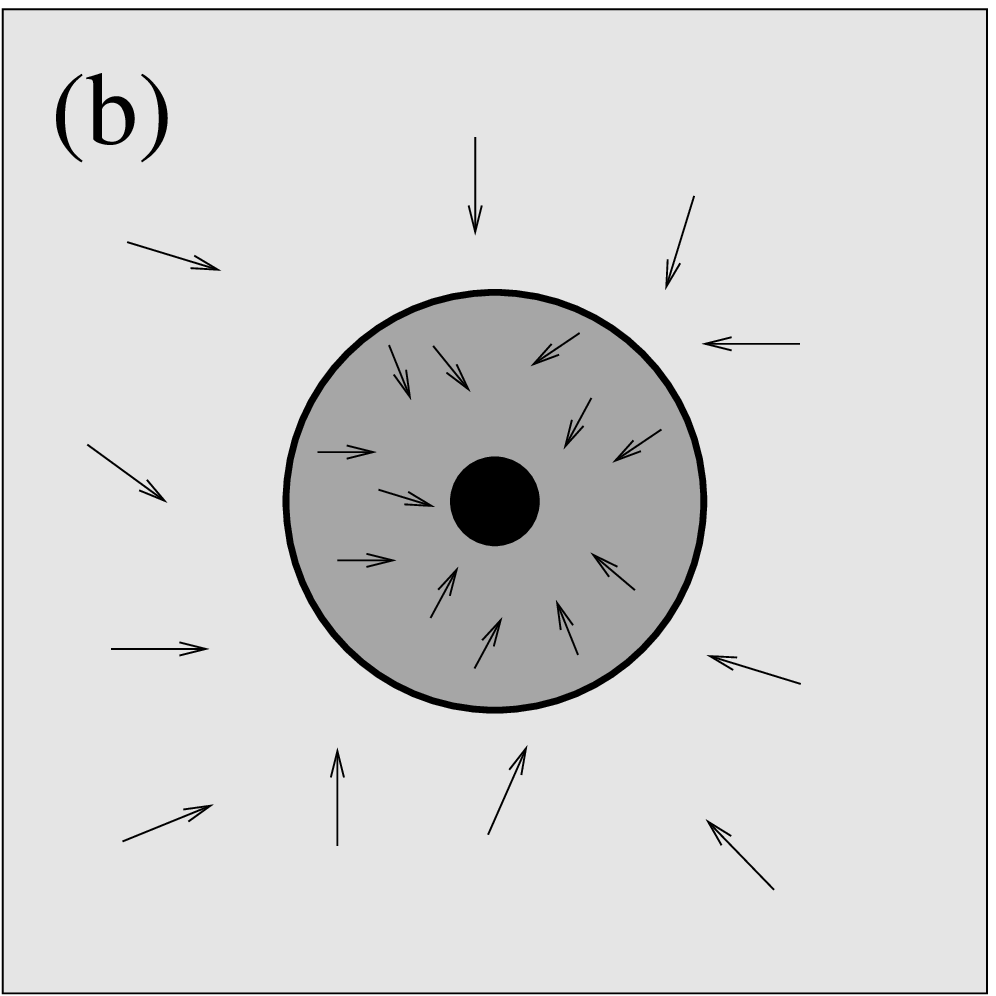,width=.5\linewidth}}
\caption{\label{fdyson}
Scheme of the Dyson sphere (a) and of its black-hole version (b). In the original version\cite{Dyson} the shell captures radiation emitted by the star inside and radiates waste heat to the space. In the black-hole version, the shell absorbs background radiation coming from outside and emits waste heat to the black hole inside.
}
\end{figure}

Rather than assuming a planet, one could imagine a spherical shell enclosing the black hole. In 1960, Freeman Dyson speculated about possible signatures of intelligent extraterrestrial life that would build a structure around a star to capture all of its power.\cite{Dyson} The waste thermal energy would be emitted as infrared radiation detectable by our observatories. We can turn this idea upside down: the inhabitants of the shell collect energy from the background radiation and send the waste thermal energy to the central black hole (see Fig.~\ref{fdyson}).

To explore the properties of such a scheme, we can use results of Sec. \ref{Sec:temperature-optimization}, however, we assume that the total area is not fixed, but variable --- determined by the radius of the Dyson sphere $R_{D}$. The sphere collects thermal energy from the outer area $S_{H} = 4\pi R_{D}^2$. With respect to the cold area, the situation is now simpler than in Sec. \ref{Sec:Etendue} since no light concentrators are necessary. The waste energy is radiated from the whole inner surface and the emitted photon either hits the black hole or is absorbed by another part of the inner surface. Thus, the waste energy is absorbed by the shadow of the black hole which for a distant observer looks as a sphere of radius $(\sqrt{27}/2)R_S$ (see Ref.~\onlinecite{Radius}). Therefore, $S_{C} = 108 \pi G^2 M^2/c^4$. 

The available power increases with increasing the Dyson sphere radius, but up to a limit given by the fixed area of the heat sink. To calculate the limiting power, we assume $R_D \gg R_S$ and apply the results of Subsection~\ref{Subsec-bigHot} with $S_C/S_H  \ll 1$. In this case the limiting power follows from Eq.~(\ref{Power}) as $P_{\rm max} \approx 0.1055 P_{\rm ref}$, where $P_{\rm ref} = \sigma S_{C}T_1^4$ is the power of the background radiation that would be absorbed by the black hole if there were no Dyson sphere. 

As an illustration, we first assume a black hole with the mass of Sun $M=2\times 10^{30}$ kg and temperature of the background radiation $T_1 = 2.725$\,K:  the limiting power is $P_{\rm max} \approx 250$\,W. 
A supermassive black hole of the size of $\sim 4\times 10^6$ solar masses (e.g., the one in the center of our Galaxy) could give us $P_{\rm max}\sim 4\times 10^{15}$\,W, i.e., about 200 times our present world energy consumption (ignoring, however, the solar power harvested by the ecosystems).
As the last example we consider the early Universe background radiation with $T_1 \approx 300$\,K and a black hole with the apparent radius equal to the radius of Sun, i.e., $(\sqrt{27}/2)R_{S} \approx 6.96\times 10^{8}$\,m. This would lead to $P_{\rm max} \approx 2.9\times 10^{20}$\,W, which is three orders of magnitude above the present solar energy income of Earth.

\section{Conclusion}
\label{Sec:Conclusion}
For nonrotating black holes and the present temperature of the cosmic microwave background, the available power appears to be rather small for the living standards of our civilization. One might speculate of the distant future when hydrogen as the nuclear fuel for stars is exhausted and black holes together with background radiation become one of the few relevant sources of negative entropy. Nevertheless, with the accelerated expansion of the Universe the background radiation becomes colder so that even less power would be available. One might also speculate of hypothetical Earth-like planets orbiting primordial black holes in the early stages of the Universe filled with room-temperature background radiation. The available power budget becomes much more generous, however, one could hardly expect that organisms with necessary radiation-focusing equipment would stand a chance to evolve.  

The situation is different for fast rotating Kerr black holes and planets on close orbits: gravitational and Doppler shifts change the temperature map of the sky to allow for harvesting much more power. Although the highly inspiring idea of the movie {\em Interstellar} to explore planets orbiting Kerr black holes is appealing, the conditions on Miller's planet of the film prove to be rather harsh. This could be expected: since the time dilation on the planet is about sixty thousand, the astronauts would receive signals from the distant outside arriving about sixty thousand times faster than emitted. Such a frequency shift must apply also to the cosmic background, making it  much hotter. Nevertheless, with a suitably chosen orbit slightly farther from Gargantua, one can hope to find the sky conditions of the planet much closer to terrestrial.


\appendix
\section{Frequency shift and temperature map of the sky for an observer orbiting a Schwarzschild black hole}
\label{Sec:analyticschw}
Here we derive the angular dependence of the frequency shift for an observer at a circular orbit around a Schwarzchild black hole.
The frequency ratio between the locally observed and emitted frequencies of ray bundles $g=\nu_{\rm obs}/\nu_\infty=p^{\langle t \rangle }/p_t$  also corresponds to the ratio of the locally measured and emitted energies (time-components of photon four-momentum). Here and hereafter, the angle brackets in the index denote the local frame of the observer on the Keplerian orbit. The source intensity divided by the third power of the frequency is conserved as a Lorentz invariant.\cite{DopplerIntensity} Moreover, Planck's law describes the black body radiation as having a spectral intensity in frequency proportional to $\nu^3 / (e^{h\nu/kT}-1)$, where $T$ is the source temperature, $\nu$ the frequency, $h$ the Planck constant and $k$ is the Boltzmann constant. Therefore, the black body spectrum of the background radiation will be locally observed as a black body spectrum with a temperature multiplied by the factor $g$ related to a particular ray bundle. Naturally, the bolometric intensity amplification of ray bundles coming from distant universe is given by the fourth power of $g$. 
Here, as usual, the bolometric intensity corresponds to the the total flux of radiation at all wavelengths per the solid angle (in W$\cdot$m$^{-2}\cdot$sr$^{-1}$).

Using the $(-+++)$ signature, geometrical units  ($c = G = 1$), and common Schwarzchild coordinates $(t,r,\vartheta,\varphi)$, the spacetime metric can be expressed in the  well-known form
\begin{align}
{\rm d}s^2 = &-\left(1-\frac{2M}{r}\right)c^2{\rm d}t^2+\frac{{\rm d}r^2}{1-\dfrac{2M}{r}} + 
\nonumber \\
& + r^2\left({\rm d}\vartheta^2+\sin^2\vartheta{\rm d}\varphi^2\right).
\end{align}
As derived in various textbooks (see, e.g. Ref.~\onlinecite{Hartle:2003}, p.~200 or Ref.~\onlinecite{Raine:2009}, p.~29), the four-velocity of an observer in the equatorial plane $\vartheta=\pi/2$ with constant $r$   has two non-zero components 
\begin{align}
u_{\rm obs}^{t}&=\frac{{\rm d}t}{{\rm d}\tau},
 \\
u_{\rm obs}^{\varphi}&=\frac{{\rm d}\varphi}{{\rm d}\tau} = \frac{{\rm d}\varphi}{{\rm d}t}\frac{{\rm d}t}{{\rm d}\tau} = u_{\rm obs}^{t}\frac{{\rm d}\varphi}{{\rm d}t},
\end{align} 
where $\tau$ is the observer's proper time. Moreover, the circular trajectories obey the Kepler's law (again, see, e.g. Refs.~\onlinecite{Raine:2009},\onlinecite{Hartle:2003})
\begin{eqnarray}
\Omega = \frac{{\rm d}\varphi}{{\rm d}t} = \sqrt{\frac{M}{r^3}}.
\end{eqnarray} 
The normalization condition ${\bf u}_{\rm obs}\boldsymbol{\cdot}{\bf u}_{\rm obs}=-1$ then leads to
\begin{eqnarray}
\frac{{\rm d}t}{{\rm d}\tau} = \frac{1}{\sqrt{1-3M/r}}
\end{eqnarray} 
and, consequently
\begin{equation}
u_{\rm obs}^{t} = \frac{1}{\sqrt{1-3M/r}},\quad
u_{\rm obs}^{\varphi} = \sqrt{\frac{M/r^3}{1-3M/r}}.  
\label{eq:fourvelobs}  
\end{equation}

Analogically\cite{Raine:2009,Hartle:2003}, for a stationary observer hovering at the same constant $r$ the only non-zero four-velocity component reads as
\begin{equation}
u_{\rm stat}^{t} = \frac{1}{\sqrt{1-2M/r}}.
  \label{eq:fourvelstat}
\end{equation}
The orbital speed $v$ of an orbiting observer with respect the stationary one at the same $r$ is determined by the Lorentz $\gamma$-factor $\gamma=\left(1-v^2\right)^{-1/2}$:
\begin{align}
\gamma &= -{\bf u}_{\rm obs}\boldsymbol{\cdot}{\bf u}_{\rm stat} = -g_{tt}u_{\rm obs}^{t}u_{\rm stat}^{t} =
\sqrt{\frac{1-2M/r}{1-3M/r}}, \\
v & = \sqrt{\frac{M/r}{1-2M/r}}.
\end{align}   
For an observer following the innermost stable circular orbit at $r=6M$ we get $v=1/2$.

The frequency shift of a photon coming from infinity includes $r$-dependent gravitational blue-shift (that would be detected also by a stationary observer) and a kinematic Doppler shift determined by the orbital velocity $v$ and the direction of light propagation. The gravitational blue-shift   is described by the well-known formula\cite{Raine:2009,Hartle:2003}
\begin{eqnarray}
\nu_{\rm stat}=\frac{\nu_{\infty}}{\sqrt{1-2M/r}},
\label{nu-stat1}
\end{eqnarray}  
where $\nu_{\infty}$ is the photon frequency in infinity and $\nu_{\rm stat}$ is the frequency detected by the stationary observer at $r$.
From the Doppler-shift formula\cite{Raine:2009,Hartle:2003} we obtain
\begin{eqnarray}
\nu_{\rm stat}=\gamma\nu_{\rm obs}\left(1+ v\cos\alpha'\right),
\label{nu-stat2}
\end{eqnarray}  
where $\alpha'$ is an angle between the observer's velocity and the direction of light propagation measured in the observer's frame of reference. Combining Eqs. (\ref{nu-stat1}) and  (\ref{nu-stat2}) we come to the final formula for the frequency shift detected by an orbiting observer
\begin{equation}
\hspace*{-2ex}g=\frac{\nu_{\rm obs}}{\nu_{\infty}}=\frac{\sqrt{1-3M/r}}{1-2M/r}\frac{1}{1+\sqrt{\dfrac{M/r}{1-2M/r}}\cos\alpha'}.  
\label{eq:shift}  
\end{equation}
The case of the innermost stable trajectory with $r=6M$ is shown in Fig.~\ref{fig:obloha}(a).
Using in Eq. (\ref{eq:shift}) $r=6M$ and  $\cos\alpha'=\mp1$, we obtain 
$g_{-} = 3/\sqrt{2} \approx2.12$ for photons striking the observer at the innermost stable trajectory from ahead (i.e., the value mentioned in Sec. \ref{sec:schwblackh}),  and $g_{+} = 1/\sqrt{2} \approx 0.71$ for photons arriving at the same observer from the rear.

Alternatively, the result can be derived from the photon four-momentum ${\bf p}_{\rm ph}$. The photon energy detected by an observer with a four-velocity ${\bf u}_{\rm obs}$ reads as\cite{Hartle:2003}
\begin{equation}
E_{\rm obs}=h\nu_{\rm obs}=-{\bf p}_{\rm ph}\boldsymbol{\cdot}{\bf u}_{\rm obs}.
  \label{eq:photen}  
\end{equation}
Because of the symmetries of the Schwarzschild metric, two quantities are conserved along light ray orbits in the equatorial plane, namely
\begin{align}
p_{\varphi}&=L_{\rm ph} = r^2p_{\rm ph}^{\varphi},\\  
p_{t}&=E_{\rm ph} = h\nu_{\infty}=\left(1-\frac{2M}{r}\right)p_{\rm ph}^{t} .
\end{align}
The quantities are connected with the photon's angular momentum $L_{\rm ph}$ and energy $E_{\rm ph}$ at infinity. From the normalization condition ${\bf p}_{\rm ph}\boldsymbol{\cdot}{\bf p}_{\rm ph}=0$ it follows
\begin{equation}
\hspace*{-1ex}p_{\rm ph}^{t} = \frac{h\nu_{\infty}}{1-2M/r},\
p_{\rm ph}^{\varphi} = \frac{h\nu_{\infty}}{r\sqrt{1-2M/r}}.
\label{eq:photonp}  
\end{equation}
For the photon in the equatorial plane these are the only non-zero four-momentum components. Because of the space-time symmetry, for a photon coming at an angle $\alpha$  seen by the stationary observer at constant $r$, substituting from~Eqs.~(\ref{eq:photonp}) and (\ref{eq:fourvelobs}) into (\ref{eq:photen}) the photon momentum can be written in the form
\begin{equation}
h\nu_{\rm obs}=-g_{tt}p_{\rm ph}^{t}u_{\rm obs}^{t}-g_{\varphi\varphi}p_{\rm ph}^{\varphi}u_{\rm obs}^{\varphi}\cos\alpha.
\end{equation}
One can express this result by means of $\alpha'$ measured by the orbiting observer on applying the relativistic aberration formula (see, e.g. Ref.~\onlinecite{Hartle:2003}, p.~93)
\begin{equation}
\cos\alpha=\frac{\cos\alpha'+v}{1+v\cos\alpha'}.
\end{equation}
After some algebraic manipulation we again come to Eq.~(\ref{eq:shift}). 
 
  
\section{Frequency shift for an observer orbiting a Kerr  black hole}
\label{Sec:Computation}

The relativistic frequency-ratio factor $g$ for an observer orbiting  a Kerr black hole is 
following the same logic as in the preceding section, but we find it numerically by computing multiple relativistic projections of ray bundles on the observer's celestial sphere.\cite{ThorneLensing,bakalaetal2015} 
The line element of the Kerr spacetime in Boyer-Lindquist coordinates parameterized by specific angular momentum (spin) $a$ reads
\begin{multline}
\label{kerr_metric}
 \mathrm{d}s^2 = -\left(1-\frac{2  r}{\Sigma}\right) \,\mathrm{d}t^2 
  - \frac{4 r a }{\Sigma} \sin^2\theta\,\mathrm{d}t\, \mathrm{d}\varphi 
  + \frac{\Sigma}{\Delta}\, \mathrm{d}r^2 + \\
+ \Sigma \,\mathrm{d}\theta^2
  + \left(r^2 + a^2 +\frac{2 r a^2 \sin^2\theta}{\Sigma}\right)\sin^2\theta\, \mathrm{d}\varphi^2, 
        \end{multline} 
where $\Sigma \equiv r^{2} + a^{2}\cos^{2}\theta$ and $\Delta \equiv r^{2} - 2r + a^{2}$. 
Components of the four-momentum of a photon in the Kerr spacetime are given by  
\begin{align}
     p^{r} &= \dot{r} = s_r\Sigma^{-1} \sqrt{R_{\lambda,q}(r)}\,,\label{CarterEQs}\\
     p^{\theta} &= \dot{\theta} = s_{\theta}\Sigma^{-1} \sqrt{\Theta_{\lambda,q}(\theta)}\,,\nonumber\\
     p^{\phi} &= \dot{\phi} = \Sigma^{-1} \Delta^{-1}
     \left[ 2ar + \lambda \left( \Sigma^{2} - 2r \right)  
\mathrm{cosec}^{2} \theta \right]\,,\nonumber\\
     p^{t} &= \dot{t} = \Sigma^{-1} \Delta^{-1} \left( \Sigma^{2} -
     2ar \lambda \right)\,,\nonumber
\end{align} 
where the dotted quantities denote differentiation with respect to some affine parameter, and the sign pair $s_{r}$,$s_{\theta}$ describes the orientation of radial and latitudinal evolution, respectively.\cite{Carter,MTW,Chandra}
Radial and latitudinal effective potentials read as 
\begin{align}
       R_{\lambda,q} \left( r \right) &=  \left( r^{2} + a^{2}
 - a \lambda \right) ^{2}
       - \Delta \left[ q + \left( \lambda - a \right) ^{2} \right]\,,\nonumber \\
       \Theta_{\lambda,q} \left( \theta \right) &= q + a^{2} \cos^{2}
\theta -\lambda^{2} \mathrm{cot}^{2} \theta\,.\label{eqn:2.1.3}
\end{align} 
Here, $\lambda$ and $ q$ are constants of motion related to the covariant components of the photon four-momentum (\ref{st.for.mom}) by the relations
\begin{equation} \label{constant_def}
  \begin{split}
\lambda &= -\frac{p_{\phi}}{p_{t}}, \,\\
q &= \left(\frac{p_{\theta}}{p_{t}}\right)^{2} + \left(\lambda\tan\left(\pi/2 - \theta\right)\right)^{2} - a^{2}\cos^{2}\theta\,.    
  \end{split}
\end{equation}
In the local reference frame related to an arbitrary observer, the azimuthal and latitudinal components $p_{\langle\varphi\rangle},\,p_{\langle\theta\rangle}$ of the photon four-momentum fully determine a projection of the corresponding ray onto the local sky. Assuming the photon energy is normalized to one, the remaining components of $p_{\langle\mu\rangle}$ can be written as follows,
\begin{equation}
\label{for.mom}
p_{\langle t \rangle} = -1, \,\,\, p_{\langle r \rangle} = \sqrt{1 - p_{\langle\theta\rangle}^{2} -  p_{\langle\varphi\rangle}^{2}}\,.
\end{equation}
One can obtain the coordinate covariant components of the four-momentum and related constants of motion (\ref{constant_def}) by transforming the local components of $p_{\langle\mu\rangle}$, using appropriate frame tetrads of one-form by the relation
\begin{equation}\label{st.for.mom}
p_{\mu} = \omega_{\mu}^{\langle\alpha\rangle}p_{\langle\alpha\rangle}.
\end{equation}
Then the frequency-ratio factor $g(\lambda,q)$ can be expressed as a function of constants of motion corresponding to the projection position on the observer's sky as
\begin{equation} \label{g-factor}
g(\lambda,q) =-\frac{1}{p_{t}(\lambda,q)}. 
\end{equation}

The local tetrad of one-forms related to the observer on the corotating Keplerian orbit around a Kerr black hole can be obtained by the Lorentz boost of ZAMO (zero angular momentum observers, locally non-rotating observers)\cite{MTW,ZAMO} tetrad, which in the equatorial plane $(\theta=\pi/2)$ takes the form \cite{ScheeStu}
\begin{equation}\label{ZAMO}
  \begin{split}
	\omega^{(t)}&=\left\{\sqrt{\frac{\Delta\Sigma}{A}},0,0,0 \right\},\\
	\omega^{(r)}&=\left\{0,\sqrt{\Sigma/\Delta},0,0 \right\},\\  
	\omega^{(\theta)}&=\left\{0,0,\sqrt{\Sigma},0 \right\},\\ 
	\omega^{(\varphi)}&=\left\{-\frac{2ar}{\sqrt{A \Sigma}},0,0,\sqrt{\frac{A}{\Sigma}}\right\},     
  \end{split}
\end{equation}
where $A \equiv (r^2 + a^2)^2 - a^2\Delta$. The velocity $\beta$ of corotating Keplerian observers with respect to the equatorial ZAMO frame can be written as \cite{ScheeStu} 
\begin{equation}
\beta=\frac{r^2+a^2-2a\sqrt{r}}{\sqrt{\Delta}(r^{3/2}+a)}.
\end{equation}
The tetrad (\ref{ZAMO}) straightforwardly transformed into the frame of corotating Keplerian observer reads
\begin{equation}
  \begin{split}
	\omega^{\langle t \rangle}&=\gamma \left\{\omega_{\,t}^{(t)}-\beta\omega^{(\varphi)}_{\,t},0,0,-\beta\omega^{(\varphi)}_{\,\varphi} \right\},\\
	\omega^{\langle r \rangle}&=\omega^{(r)}, \qquad \omega^{\langle\theta\rangle}=\omega^{(\theta)},\\ 
	\omega^{\langle\varphi\rangle}&=\gamma \left\{\omega^{(\varphi)}_{\,t}-\beta\omega^{(t)}_{\,t},0,0,\omega^{(\varphi)}_{\,\varphi} \right\},   
  \end{split}
\end{equation}
where $\gamma \equiv (1-\beta^2)^{-1/2}$.

For such an observer located in the close vicinity of the event horizon, the factor $g$ varies depending on angular coordinates in the local sky, as a result of interplay between extreme gravitational lensing and optical effects of special relativity caused by the orbital motion of the Keplerian frame. In principle, applying the relations discussed above, it is possible to construct the so-called critical loci curve, which forms a boundary between a projection of a distant universe and a shadow of the black hole in the local sky.\cite{Viergutz,Teo} To avoid significant analytical difficulties associated with such an approach, we used our relativistic ray-tracing code LSDplus, which performs a time-reverse direct numerical integration of the equations~(\ref{CarterEQs})\cite{bakalaetal2007,bakalaetal2015}. 

The projection of black hole shadow on the local celestial sphere corresponds to rays coming to the observer from a very close vicinity of the event horizon. The most blue-shifted rays coming from distant universe are projected near the right side of the shadow edge and around the celestial equator (see Fig.~\ref{fig:obloha}). The shadow edge corresponds to rays tightly passing or just leaving the boundary of black hole photosphere.\cite{Photosphere} Therefore the ray-tracing code must integrate photon trajectories as accurately as possible in order to precisely distinguish between the extremely blue-shifted rays coming from infinity and the rays leaving the photosphere. The Runge-Kutta method of the eighth order (Dorman-Prince method) \citep{NR} used here integrates the null geodesics with the very satisfactory relative accuracy of $10^{-15}$, which, in the case of a central black hole with stellar mass, corresponds to the order of accuracy of $10^{-11}$ metres in the radial coordinate \cite{bakalaetal2015}. Moreover, the code LSDplus allows one to focus on arbitrarily chosen and arbitrarily sized rectangular part of the local sky in the Mollweide projection (see Fig.~\ref{fig:vecko1}) with resolution limited only by the size of computer memory. We have used the screen resolution of 4000 x 2000 pixels.

The techniques described above allow one to trace the location of the shadow edge with a high precision (see Figs~\ref{fig:obloha}, \ref{fig:vecko1}). The relativistic frequency-ratio factor $g$ given by (\ref{g-factor}) for ray bundles coming from distant universe is calculated by the code as well.



\begin{thebibliography}{99}
\bibitem{EarthEnergy}
The energy input/output of the Earth is the solar constant $\sim$ 1.36\,kW/m$^2$ multiplied by the disk area of the Earth $\pi R^2 \approx 1.28 \times 10^{14}$\,m$^2$, i.e., 1.74$\times 10^{17}$\,J each second; if this is divided by the temperature $\sim$\,300\,K we obtain the estimate of the entropy production.
\bibitem{Celestial}
Celestial sphere is an imaginary sphere of arbitrarily large radius concentric with the observer. It is a practical tool in spherical astronomy for projecting objects on the sky. In Fig.~\ref{fig:obloha} we show the Mollweide projection of the celestial sphere with the light (dark) area representing its hot (cold) part, respectively. 
The Mollweide projection generally used for global maps of the world or night sky is an equal-area, pseudocylindrical map projection preserving the proportions of areas, which is important for our considerations. It represents the central meridian as a straight line segment with other meridians longer and bowed outward away from the central one. The projection maps the sphere into a proportional 2:1 ellipse (the length of the central meridian is a half of the equator length).
\bibitem{Spin}
The rotation parameter $a$ (spin) is defined as a specific intrinsic angular momentum of the Kerr black hole \cite{MTW, Chandra}. For $a=0$ the Kerr solution merges into Schwarzschild non-rotating spherically symmetric spacetime (see eq.~(\ref{kerr_metric})). Maximum possible ``black hole'' value $a=1$ corresponds to an extreme Kerr black hole having the rotation velocity of the event horizon equal to the speed of light. The case of hypothetical naked singularities is described by $a>1$.
\bibitem{Photosphere}
Kerr black hole photosphere -- region of unstable spherical or circular photon orbits wrapping the black hole. The  photosphere reaches its maximum extent in the equatorial plane between corotating and counter-rotating circular photon orbits, while it becomes infinitesimally thin on the polar axis.\cite{Teo}
\end{thebibliography}
\end{document}